\documentclass[letterpaper,12pt]{article}
\usepackage{osajnl2_arxiv}
\usepackage{textcomp,amsfonts}

\newcommand{\micron}{\textmu m}

\begin{document}

\title{Shape reconstruction from gradient data}

\author{Svenja~Ettl,\footnote{n\'{e}e Lowitzsch}$^{,*}$ J\"urgen~Kaminski, Markus~C.~Knauer, and Gerd~H{\"a}usler}

\address{Institute of Optics, Information and Photonics, Max Planck Research Group,
\\University Erlangen-Nuremberg, Staudtstr. 7/B2, 91058 Erlangen, Germany}

\address{$^*$Corresponding author: settl@optik.uni-erlangen.de}

\begin{abstract}
We present a novel method for reconstructing the shape of an object from measured gradient data. A certain class of optical sensors does not measure the shape of an object, but its local slope. These sensors display several advantages, including high information efficiency, sensitivity, and robustness. For many applications, however, it is necessary to acquire the shape, which must be calculated from the slopes by numerical integration. Existing integration techniques show drawbacks that render them unusable in many cases. Our method is based on approximation employing radial basis functions. It can be applied to irregularly sampled, noisy, and incomplete data, and it reconstructs surfaces both locally and globally with high accuracy.
\end{abstract}


\maketitle

\section{Introduction}

In industrial inspection, there is an ever-growing demand for highly accurate, non-destructive measurements of three-dimensional object geometries. A variety of optical sensors have been developed to meet these demands~\cite{Girod00}.
These sensors satisfy the requirements at least partially. Numerous applications, however, still wait for a capable metrology. The limitations of those sensors emerge from physics and technology--the physical limits are determined by the wave equation and by coherent noise, while the technological limits are mainly due to the space-time-bandwidth product of electronic cameras.

\begin{figure}[htbp]
\centerline{\includegraphics[width=16.6cm]{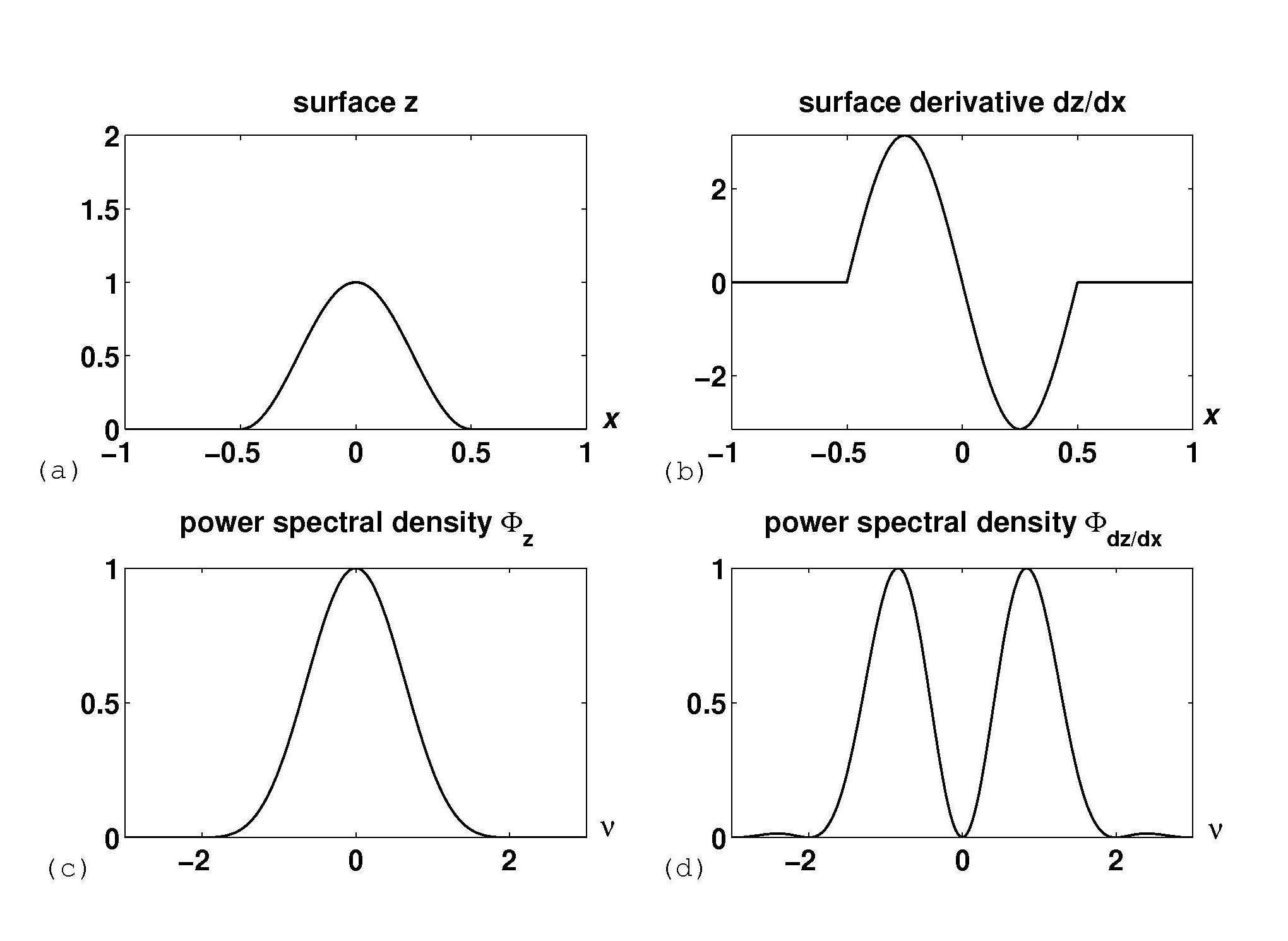}}
\caption{(a) Surface $z$ of a typical smooth object and (b) its derivative $\frac{dz}{dx}$, together with (c) the power spectral density of the surface $\Phi_z$ and (d) of its slope $\Phi_\frac{dz}{dx}$ (all in arbitrary units). The power spectral density shows that differentiation reduces redundancy, contained in the low frequencies. }
\label{spectra}
\end{figure}

Closer consideration reveals that the technological limits are basically of information-theoretical nature. The majority of the available optical 3D sensors need large amounts of raw data in order to obtain the shape. A lot of redundant information is acquired and the expensive channel capacity of the sensors is used inefficiently~\cite{Wagner03}. A major source of redundancy is the shape of the object itself: If the object surface $z(x,y)$ is almost planar, there is similar height information at each pixel. In terms of information theory the surface points of such objects are ``correlated''; their power spectral density $\Phi_z$ decreases rapidly.
In order to remove redundancy, one can apply spatial differentiation to whiten the power spectral density (see Fig.~\ref{spectra}). Fortunately, there are optical systems that perform such spatial differentiation. Indeed, sensors that acquire just the local slope instead of absolute height values are much more efficient in terms of exploiting the available channel capacity. Further, reconstructing the object height from slope data reduces the high-frequency noise since integration acts as a low-pass filter. 

There are several sensor principles that acquire the local slope: For \emph{rough surfaces}, it is mainly the principle of Shape from Shading~\cite{Horn89}. For \emph{specular surfaces}, there are differentiating sensor principles like the differential interference contrast microscopy or deflectometry~\cite{Haeusler88}. Deflectometric scanning methods allow an extremely precise characterization of optical surfaces by measuring slope variations as small as $0.02$\,arcsec~\cite{Weingaertner99}. Full-field deflectometric sensors acquire the two-dimensional local gradient of a (specular) surface. Using ``Phase-measuring Deflectometry'' (PMD)~\cite{PMD99Pt,Knauer04,Kaminski05a,Lowitzsch05b}, for example, one can measure the local gradient of an object at one million sample points within a few seconds. The repeatability of the sensor described in~\cite{Lowitzsch05b} is below $10$\,arcsec with an absolute error less than $100$\,arcsec, on a measurement field of $80$\,mm~$\times$~$80$\,mm and a sampling distance of $0.1$\,mm.

In several cases it is sufficient to know the local gradient or the local curvature; however, most applications demand the height information as well. As an example we consider eyeglass lenses. In order to calculate the local surface power of an eyeglass lens by numerical differentiation, we only need the surface slope and the lateral sampling distance. But for quality assurance in an industrial setup, it is necessary to adjust the production machines according to the measured shape deviation. This requires height information of the surface. Another application is the measurement of precision optics. For the optimization of these systems sensors are used to measure the local gradient of wavefronts~\cite{Pfund98}. To obtain the shape of these wavefronts, a numerical shape reconstruction method is needed.

\subsection{Why is integration of 2D gradient data difficult?}
In the previous section we stated that measuring the gradient instead of the object height is more efficient from an information-theoretical point of view, since redundant information is largely reduced. Using numerical integration techniques, the shape of the object can be reconstructed locally with high accuracy. For example, a full-field deflectometric sensor allows the detection of local height variations as small as \emph{a few nanometers}.

However, if we want to reconstruct the global shape of the object, low-frequency information is essential. Acquiring \emph{solely the slope\/} of the object reduces the low-frequency information substantially (see Fig.~\ref{spectra}). In other words, we have a lot of \emph{local\/} information while lacking \emph{global\/} information, because we reduced the latter by optical differentiation. As a consequence, small measuring errors in the low-frequency range will have a strong effect on the overall reconstructed surface shape. This makes the reconstruction of the global shape a difficult task. 

Furthermore, one-dimensional integration techniques cannot be easily extended to the two-dimensional case. In this case, one has to choose a path of integration. Unfortunately, noisy data leads to different integration results depending on the path\cite{Agrawal05}. Therefore, requiring the integration to be path independent becomes an important condition ("integrability condition") for developing an optimal reconstruction algorithm (see Sections~\ref{problem} and~\ref{optimal}).

\section{Problem formulation}
\label{problem}

We consider an \emph{object surface\/} to be a twice continuously differentiable function $z: \Omega\to \mathbb{R}$ on some compact, simply connected region $\Omega\subset \mathbb{R}^2$. The integrability condition implies that the gradient field $\nabla z=(z_x,z_y)^T$ is \emph{curl free}, i.\,e. every path integral between two points yields the same value. This is equivalent to the requirement that there exists a potential $z$ to the gradient field $\nabla z$ which is unique up to a constant. Most object surfaces measurable by deflectometric sensors fulfill these requirements, or at least they can be decomposed into simple surface patches showing these properties. 

Measuring the gradient $\nabla z$ at each sample point $\mathbf{x}_i=(x_i,y_i)^T$ yields a discrete vector field $(p(\mathbf{x}_i),q(\mathbf{x}_i))^T,\; i=1\dots N$. These measured gradient values usually are contaminated by noise--the vector field is not necessarily curl free. Hence, there might not exist a potential $z$ such that $\nabla z(\mathbf{x}_i) = (p(\mathbf{x}_i),q(\mathbf{x}_i))^T$ for all $i$. In that case, we seek a least-squares approximation, i.\,e.\ a surface representation $z$ such that the following error functional is minimized~\cite{Horn89,Frankot88,HorovitzK04}:
\begin{equation}\label{approx_functional}
	J(z) := \sum_{i=1}^N \left[z_x({\bf x}_i)-p({\bf x}_i)\right]^2 + \left[z_y({\bf x}_i)-q({\bf x}_i)\right]^2.
\end{equation}

\section{Related Work}

In the case of one-dimensional data, integration is a rather straightforward procedure which has been investigated quite extensively~\cite{Yaroslavsky,Elster02a}. In case of two-dimensional data, there exist mainly two different approaches to solve the stated problem~\cite{Schluens96}. 
On the one hand, there are \emph{local methods}: They integrate along predetermined paths. The advantage of these methods is that they are simple and fast, and that they reconstruct small local height variations quite well. However, they propagate both the measurement error and the discretization error along the path. Therefore, they introduce a global shape deviation. This effect is even worse if the given gradient field is not guaranteed to be curl free: In this case, the error also depends on the chosen path.
On the other hand, there are \emph{global methods}: They try to minimize $J(z)$ by solving its corresponding Euler-Lagrange equation\cite{Agrawal05}
\begin{equation}\label{poisson}
	\nabla^2 z({\bf x}) = \frac{\partial p}{\partial x}({\bf x}) + \frac{\partial q}{\partial y}({\bf x})\,,
\end{equation}
where $\frac{\partial p}{\partial x}$ and $\frac{\partial q}{\partial y}$ denote the numerical $x$- and $y$-derivative of the measured data $p$ and $q$, respectively. The advantage of global methods is that there is no propagation of the error; instead, the error gets evenly distributed over all sample points. Unfortunately, the implementation has certain difficulties. Methods based on finite differences are usually inefficient in their convergence when applied to strongly curved surfaces~\cite{HorovitzK04}. Therefore, they are mainly used for nearly planar objects. Another approach to solve Eq.~(\ref{poisson}) is based on Fourier transformation~\cite{Frankot88}. Integration in the Fourier domain has the advantage of being optimal with respect to information-theoretical considerations~\cite{Yaroslavsky}. However, Fourier methods assume a periodic extension on the boundary which cannot be easily enforced with irregularly shaped boundaries in a two-dimensional domain.   

In general, it is crucial to note that the reconstruction method depends on the slope-measuring sensor and the properties of the acquired data. For example, slope data acquired by Shape from Shading is rather noisy, exhibits curl, and is usually located on a full grid of millions of points. Here, a fast subspace approximation method like the one proposed by Frankot and Chellappa~\cite{Frankot88} is appropriate. On the other hand, wavefront reconstruction deals with much smaller data sets, and the surface is known to be rather smooth and flat. In this case, a direct finite-difference solver can be applied~\cite{Southwell80}. Deflectometric sensors deliver a third type of data: It consists of very large data sets with rather small noise and curl, but the data may not be complete, depending on the local reflectance of the measured surface. Furthermore, the measuring field may have an unknown, irregularly shaped boundary. These properties render most of the aforementioned methods unusable for deflectometric data.
In the following sections, we will describe a surface reconstruction method which is especially able to deal with slope data acquired by sensors such as Phase-measuring Deflectometry.

\section{Shape reconstruction}
\label{method}

\subsection{Challenges}
The desired surface reconstruction method should have the properties of both \emph{local and global\/} integration methods: It needs to \emph{preserve local details\/} without propagating the error along a certain path. It also needs to minimize the error functional of Eq.~(\ref{approx_functional}), hence yielding a \emph{globally optimal solution\/} in a least-squares sense. Further, the method should be able to deal with \emph{irregularly shaped boundaries}, \emph{missing data points}, and it has to be able to reconstruct surfaces of a large variety of objects with steep slopes and \emph{high curvature values}. It should also be able to \emph{handle large data sets\/} which may consist of some million sample points. 
We now show how to meet these challenges using an analytic interpolation approach.

\subsection{Analytic reconstruction}
A low noise level allows \emph{interpolation\/} of the slope values instead of approximation. Interpolation is a special case which has the great advantage that we can ensure that small height variations are preserved. In this paper we will only focus on the interpolation approach as analytic reconstruction method. For other measurement principles like Shape from Shading, an approximation approach might be more appropriate. 

The basic idea of the integration method is as follows: We seek an analytic interpolation function such that its gradient  interpolates the measured gradient data. Once this interpolation is determined, it uniquely defines the surface reconstruction up to an integration constant. To obtain the analytic interpolant, we choose a generalized Hermite interpolation approach employing \emph{radial basis functions\/} (RBFs)~\cite{Narcowich03,Buhmann03}. This method has the advantage that it can be applied to \emph{scattered data}. It allows us to integrate data sets with holes, irregular sampling grids, or irregularly shaped boundaries. Furthermore, this method allows for an optimal surface recovery in the sense of Eq.~(\ref{approx_functional}) (see Section~\ref{optimal} below).

In more detail: Assuming that the object surface fulfills the requirements described in Section~\ref{problem}, the data is given as pairs $(p(\mathbf{x}_j),q(\mathbf{x}_j))^T$, where $p(\mathbf{x}_j)$ and $q(\mathbf{x}_j)$ are the measured slopes of the object at $\mathbf{x}_j$ in $x$- and $y$-direction, respectively, for $1 \le j \le N$. 
We define the interpolant to be
\begin{equation}\label{interpolant}
	s({\bf x}) = \sum_{i=1}^{N} \alpha_i \, \Phi_x (\mathbf{x}-\mathbf{x}_i) 
	           + \sum_{i=1}^{N} \beta_i \, \Phi_y (\mathbf{x}-\mathbf{x}_i),
\end{equation}
where $\alpha_i$ and $\beta_i$, for $1 \le i \le N$, are coefficients and $\Phi:\mathbb{R}^2 \to\mathbb{R}$ is a radial basis function. Hereby, $\Phi_x$ and $\Phi_y$ denote the analytic derivative of $\Phi$ with respect to $x$ and $y$, respectively. This interpolant is specifically tailored for gradient data~\cite{Lowitzsch05b}. To obtain the coefficients in Eq.~(\ref{interpolant}) we match the analytic derivatives of the interpolant with the measured derivatives:
\begin{equation}\label{setup_interpolation}
	\left\{
	\begin{array}{r@{\quad}}
	\displaystyle s_x (\mathbf{x}_j) \stackrel{!}{=} p(\mathbf{x}_j) \\
	\\
	\displaystyle s_y (\mathbf{x}_j) \stackrel{!}{=} q(\mathbf{x}_j) 
	\end{array}
	\right.
	\mbox{, \quad for $1 \le j \le N$}.
\end{equation}
This leads to solving the following system of linear equations~\cite{Ettl07}:
\begin{equation}
\label{linear_system}
	\underbrace{\left(
	\mbox{\begin{tabular}{cc} 
	$\Phi_{x x}(\mathbf{x}_i-\mathbf{x}_j)$ & $\Phi_{x y}(\mathbf{x}_i-\mathbf{x}_j)$\\
	\\[-0.7em]
	\\[-0.7em]
	$\Phi_{x y}(\mathbf{x}_i-\mathbf{x}_j)$ & $\Phi_{y y}(\mathbf{x}_i-\mathbf{x}_j)$\\
	\end{tabular}}
	\right)}_{A \, \in \, M^{2N\times2N}}
	\underbrace{\left(
	\mbox{\begin{tabular}{c} 
	$\alpha_i$\\
	\\[-0.7em]	
	\\[-0.7em]	
	$\beta_i$
	\end{tabular}}
	\right)}_{\alpha \, \in \, M^{2N\times1}}
	=
	\underbrace{\left(
	\mbox{\begin{tabular}{c} 
	$p(\mathbf{x}_j)$\\
		\\[-0.7em]	
	\\[-0.7em]	
	$q(\mathbf{x}_j)$
	\end{tabular}}
	\right)}_{d \, \in \, M^{2N\times1}}.
\end{equation}
\\
Using the resulting coefficients $\alpha_i, \beta_i$ we then can apply the interpolant in Eq.~(\ref{interpolant}) to reconstruct the object surface. For higher noise levels an \emph{approximation\/} approach is recommended. In this case, we simply reduce the number of basis functions so that they do not match the number of data points any more. The system $A\,\alpha=d$ in Eq.~(\ref{linear_system}) then becomes overdetermined and can be solved in a least-squares sense.

\subsection{Optimal recovery}\label{optimal}
The interpolation approach employing radial basis functions has the advantage that it yields a \emph{unique\/} solution to the surface recovery problem: Within this setup, the interpolation matrix in Eq.~(\ref{linear_system}) is always symmetric and positive definite. Further, the solution satisfies a \emph{minimization principle\/} in the sense that the resulting analytic surface function has minimal energy~\cite{DNW99}.

We choose $\Phi$ to be a Wendland's function~\cite{Wendland95}, $\Phi({\bf x}) =: \phi(r)$, with  
\begin{equation}\label{wendland}
	\phi(r) = \frac{1}{3}(1-r)_+^6(35r^2+18r+3)\in C^4(\mathbb{R}_+) \mbox{~and~} r:=\sqrt{x^2+y^2}.
\end{equation}
This has two reasons: First, Wendland's functions allow to choose their continuity according to the smoothness of the given data. The above Wendland's function leads to an interpolant which is three-times continuously differentiable, hence guaranteeing the integrability condition. Second, the compact support of the function allows to adjust the support size in such a way that the solution of Eq.~(\ref{linear_system}) is stable in the presence of noise. It turns out that the support size has to be chosen rather large in order to guarantee a good surface reconstruction~\cite{Lowitzsch05b}.

\subsection{Handling large data sets}
The amount of data commonly acquired with a PMD sensor in a single measurement is rather large: it consists of about one million sample points. This amount of data, which results from a measurement with high lateral resolution, would require the inversion of a matrix with $(2\times 10^6)^2$ entries (Eq.~(\ref{linear_system})). Since we choose a large support size for our basis functions to obtain good numerical stability the corresponding matrix is not sparse. It is obvious that this large amount of data cannot be handled directly by inexpensive computing equipment in reasonable time.

\begin{figure}[htbp]
\centerline{\includegraphics[width=16.6cm]{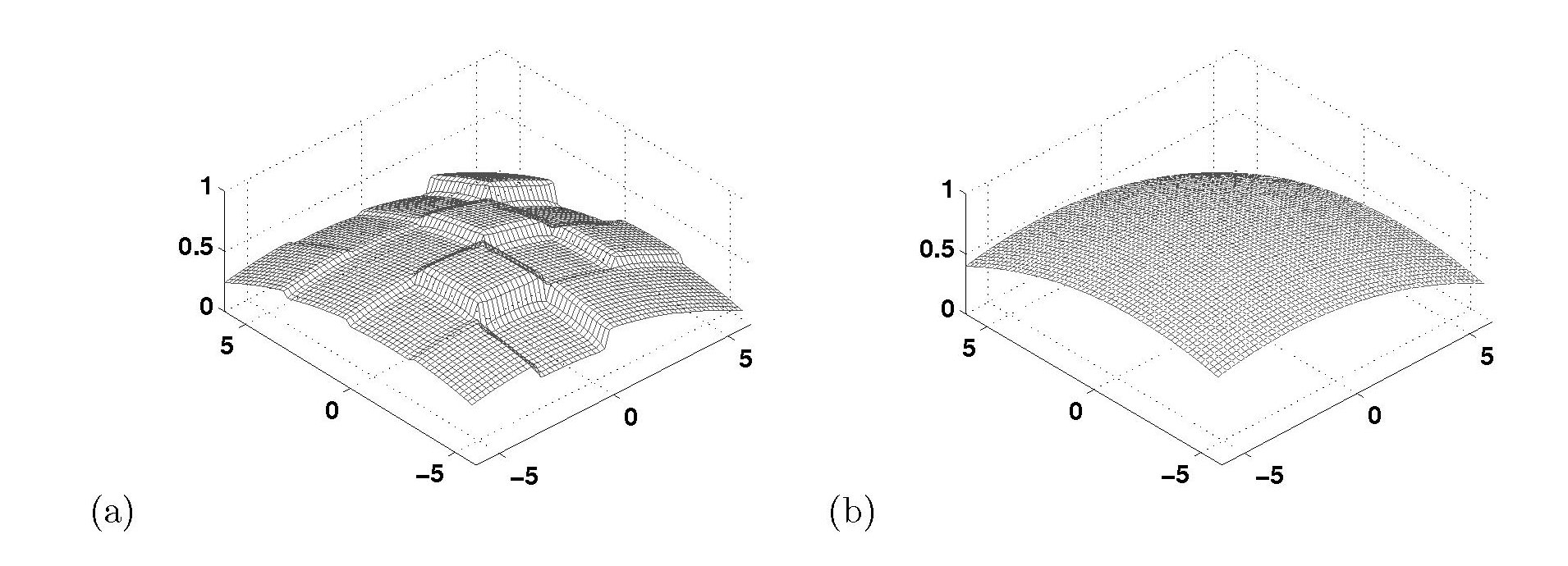}}
\caption{Reconstructed surface of a spherical lens from its local slope data (a) before and (b) after alignment of patches (all units in mm). The patches were introduced in order to handle arbitrarily large data sets. }
\label{mirror_fit}
\end{figure}

To cope with such large data sets we developed a method which first splits the data field into a set of overlapping rectangular patches. We interpolate the data on each patch separately. If the given data were height information only, this approach would yield the complete surface reconstruction. For slope data, we interpolate the data and obtain a surface reconstruction \emph{up to a constant of integration\/} (see Fig.~\ref{mirror_fit}(a)) on each patch. In order to determine the missing information we apply the following fitting scheme:
Let us denote two adjacent patches as $\Omega_1$ and $\Omega_2$ and the resulting interpolants as $s_1$ and $s_2$, respectively. Since the constant of integration is still unknown the two interpolants might be on different height levels. Generally, we seek a correcting function $f_2:\Omega_2\to\mathbb{R}$ by minimizing
\begin{equation}\label{lsqfit}
	K(f_2) := \sum_{\mathbf{x}\in{\Omega_1}\cap{\Omega_2}} |s_1(\mathbf{x}) - s_2(\mathbf{x}) - f_2(\mathbf{x})|^2\,.
\end{equation}
This fitting scheme is then propagated from the center toward the borders of the data field to obtain the reconstructed surface on the entire field (see Fig.~\ref{mirror_fit}(b)).

In the simplest case, the functions $f_i$ are chosen to be constant on each patch, representing the missing constant of integration. If the systematic error of the measured data is small, the constant fit method is appropriate since it basically yields no error propagation. For very noisy data sets it might be better to use a planar fit, i.\,e.\ $f_i(\mathbf{x}) = a_i\,x+b_i\,y+c_i$, to avoid discontinuities at the patch boundaries. This modification, however, introduces a propagation of the error along the patches. The correction angle required on each patch to minimize Eq.~(\ref{lsqfit}) depends on the noise of the data. Numerical experiments have shown that in most cases the correction angle is at least ten times smaller than the noise level of the measured data. 

Using this information we can estimate the global height error which, by error propagation, may sum up toward the borders of the measuring field~\cite{Knauer06a}:
\begin{equation}\label{globaltilt}
	\Delta z_{\rm global} \approx \tan(\sigma_\alpha)\,\Delta x\,\sqrt{M}\,,
\end{equation}
where $\sigma_\alpha$ is the standard deviation of the correction angles, $\Delta x$ is the patch size, and $M$ is the number of patches. Suppose we want to integrate over a field of $80$\,mm (which corresponds to a typical eyeglass diameter), assuming a realistic noise level of $8$\,arcsec and a patch size (not including its overlaps) of $3$\,mm. With our setup, this results in $27\times 27$ patches, with a maximal tipping of $\sigma_\alpha\approx 0.6$\,arcsec per patch. According to Eq.~(\ref{globaltilt}), the resulting global error caused by propagation of the correction angles is \emph{only $45$\,nm}. 

We choose the size of the patches as big as possible, provided that a single patch can still be handled efficiently. For the patch size in the example, $23\times 23$ points (including $25\%$ overlap) correspond to a $1058\times 1058$ interpolation matrix that can be inverted quickly using standard numerical methods like Cholesky decomposition. 

A final remark concerning the runtime complexity of the method described above: The complexity can be further reduced in case the sampling grid is regular. Since the patches all have the same size and the matrix entries in Eq.~(\ref{linear_system}) only depend on the distances between sample points, the matrix can be inverted once for all patches and then applied to varying data on different patches, as long as the particular data subset is complete.
Note that Eq.~(\ref{interpolant}) can be written as $s = B\,\alpha$, where $B$ is the evaluation matrix. Then, by applying Eq.~(\ref{linear_system}) we obtain $s = BA^{-1}\,d$, where the matrix $BA^{-1}$ needs to be calculated only once for all complete patches. If samples are missing, however, the interpolation yields different coefficients $\alpha$ and hence forces to recompute $BA^{-1}$ for this particular patch. Using these techniques, the reconstruction of $1000\times 1000$ surface values from their gradients takes about $5$ minutes on a current personal computer.

\section{Results}

\begin{figure}[htbp]
\centerline{\includegraphics[width=14.0cm]{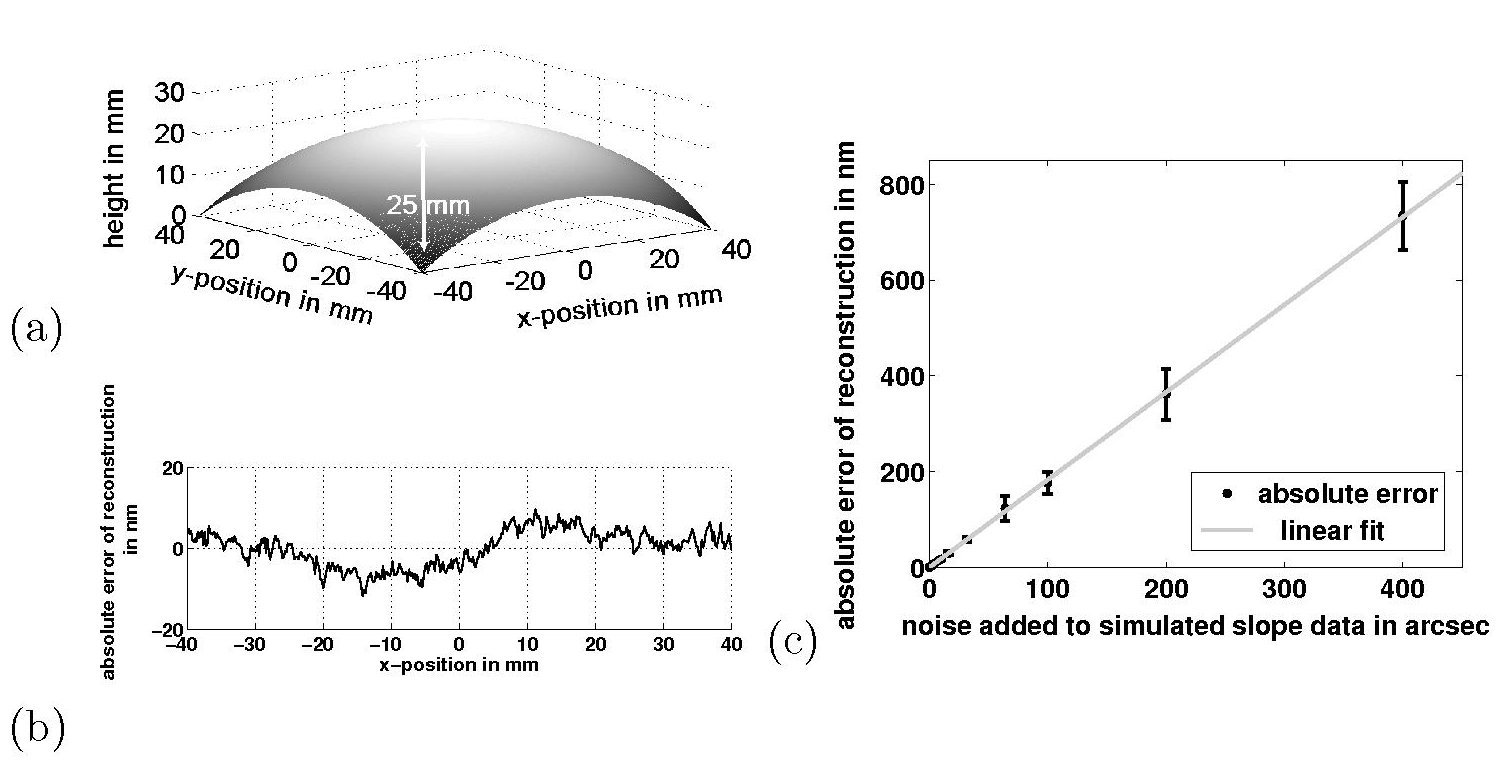}}
\caption{Surface reconstruction for simulated, noisy slope data of a sphere: (a) The reconstructed spherical surface and (b) a cross-section of the absolute error of the reconstruction, both for realistic noise of $8$\,arcsec, and (c) the absolute error of the reconstruction for several noise levels. The absolute error increases only linearly. }
\label{noise_sphere}
\end{figure}

First, we investigated the stability of our method with respect to noise. We simulated realistic gradient data of a sphere (with $80$\,mm radius, $80$\,mm~$\times$~$80$\,mm field with sampling distance $0.2$\,mm, see Figure~\ref{noise_sphere}(a)) and added uniformly distributed noise of different levels, ranging from $0.05$ to $400$\,arcsec. We reconstructed the surface of the sphere using the interpolation method described in Section~\ref{method}. Hereby, we aligned the patches by only adding a constant to each patch. The reconstruction was performed for $12$ statistically independent slope data sets for each noise level. 

Depicted in Figure~\ref{noise_sphere}(b) is a cross-section of the absolute error of the surface reconstruction from the ideal sphere, for a realistic noise level of $8$\,arcsec. The absolute error is less than $\pm 15$\,nm \emph{on the entire measurement field}. The local height error corresponding to this noise level is about $\pm 5$\,nm. This demonstrates that the dynamic range of the global absolute error with respect to the height variance ($25$\,mm) of the considered sphere is about $1:10^6$ . 

The graph in Figure~\ref{noise_sphere}(c) depicts the mean value and the standard deviation (black error bars) of the absolute error of the reconstruction, for $12$ different data sets and for different noise levels. It demonstrates that for increasing noise level the absolute error grows only linearly (linear fit depicted in gray), and even for a noise level being the \emph{fifty-fold of the typical sensor noise\/} the global absolute error remains in the \emph{sub-micrometer regime}. This result implies that the reconstruction error is smaller than most technical applications require.

\begin{figure}[htbp]
\centerline{\includegraphics[width=14.0cm]{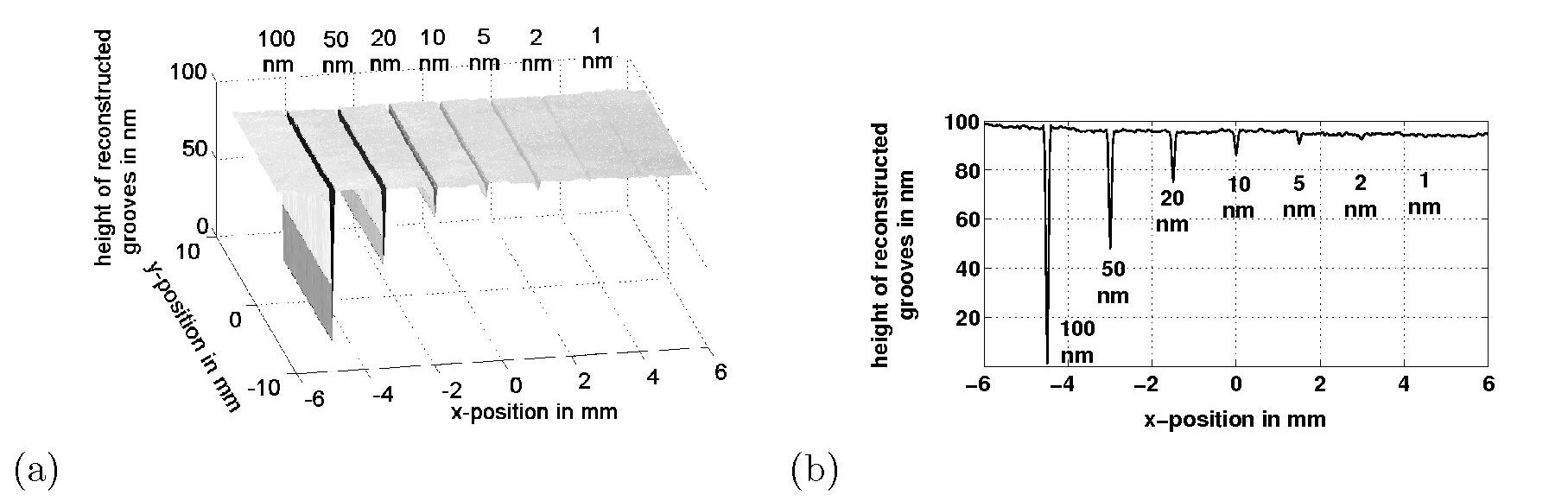}}
\caption{Reconstruction of grooves on a spherical surface from simulated slope data, for realistic noise. The nominal height of the grooves ranges from $100$\,nm~down to $1$\,nm. After the reconstruction, the sphere was subtracted to make the grooves visible. The actual, reconstructed grooves are depicted in (a) full-field and in (b) cross-section.}
\label{groove_sphere}
\end{figure}

Another common task in quality assurance is the detection and quantification of surface defects like scratches or grooves. We therefore tested our method for its ability to reconstruct such local defects that may be in a range of only a few nanometers. For this purpose, we considered data from a PMD sensor for small, specular objects. The sensor has a resolvable distance of $30$\,\micron\ laterally and a local angular uncertainty of about $12$\,arcsec~\cite{Kaminski05a}. In order to quantify the deviation of the perfect shape, we again simulated a sphere (this time with $12$\,mm radius and $5.7$\,mm~$\times$~$5.7$\,mm data field size). We added parallel, straight grooves of varying depths from $1$ to $100$\,nm and of $180$\,\micron\ width and reconstructed the surface from the modified gradients. The perfect sphere was then subtracted from the reconstructed surface. 
The resulting reconstructed grooves are depicted in Figure~\ref{groove_sphere}(a). The grooves ranging from $100$ down to $5$\,nm depth are clearly distinguishable from the plane. Figure~\ref{groove_sphere}(b) shows that all reconstructed depths agree fairly well with the actual depths. Note that each groove is determined by only $5$ inner sample points. The simulation results demonstrate that our method is almost free of error propagation while preserving small, local details of only some nanometers height. 

\begin{figure}[htbp]
\centerline{\includegraphics[width=12.0cm]{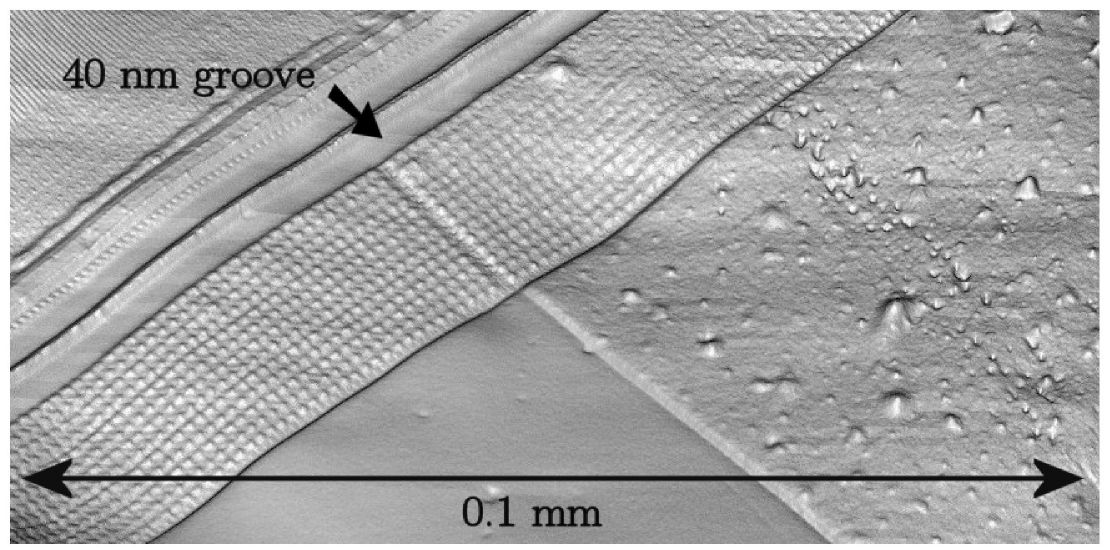}}
\caption{Height reconstruction of a part of a wafer from its local slope data. The data was acquired by a Phase-measuring Deflectometry sensor (measurement field $100$\,\micron$ \times 80$\,\micron, height range about $350$\,nm). The diagonal groove (marked by the arrow) has a depth of about $40$\,nm. }
\label{wafer}
\end{figure}

So far, we used only simulated data to test the reconstruction. Now, we want to demonstrate the application of our method to a real measurement. The measurement was performed with a Phase-measuring Deflectometry sensor for very small, specular objects. It can laterally resolve object points with a distance of $75$\,nm, while having a local angular uncertainty of about $200$\,arcsec. The object under test is a part of a wafer with about $350$\,nm height range. The size of the measurement field was $100$\,\micron~$\times$~$80$\,\micron. Depicted in Figure~\ref{wafer} is the reconstructed object surface from roughly three million data values. Both the global shape and local details could be reconstructed with high precision.

\section{Conclusion}

We motivated why the employment of optical slope-measuring sensors can be advantageous. We gave a brief overview of existing sensor principles. The question that arose next was how to reconstruct the surface from its slope data. We presented a method based on radial basis functions which enables us to reconstruct the object surface from noisy gradient data. The method can handle large data sets consisting of some million entries. Furthermore, the data does not need to be acquired on a regular grid--it can be arbitrarily scattered and it can contain data holes. We demonstrated that, while accurately reconstructing the object's global shape, which may have a \emph{height range of some millimeters}, the method preserves \emph{local height variations on a nanometer scale}. 

A remaining challenge is to improve the runtime complexity of the algorithm in order to be able to employ it for inline quality assurance in a production process.

\section*{Acknowledgments}

The authors thank the Deutsche Forschungsgemeinschaft for supporting this work in the framework of SFB 603.



\newpage

\newpage

\newpage

\newpage

\newpage

\end{document}